\documentclass[journal=jpclcd,manuscript=letter]{achemso}
\SectionNumbersOn
\usepackage{setspace}
\usepackage{amsmath,amssymb}
\usepackage{graphicx}
\usepackage{bm}
\usepackage{multirow}
\usepackage{caption}
\usepackage{cases}
\title{Steady State Photoisomerization Quantum Yield of Model Rhodopsin: Insights from Wavepacket Dynamics?}
\author{Chern Chuang}
\email{chern.chuang@utoronto.ca}
\author{Paul Brumer}
\email{Paul.Brumer@utoronto.ca}
\affiliation{Chemical Physics Theory Group, Department of Chemistry, and Center for Quantum Information and Quantum
Control, University of Toronto, Toronto, Ontario M5S 3H6, Canada}
\begin{document}
\setlength{\fboxrule}{0 pt}

\begin{tocentry}
  \includegraphics[width=10cm]{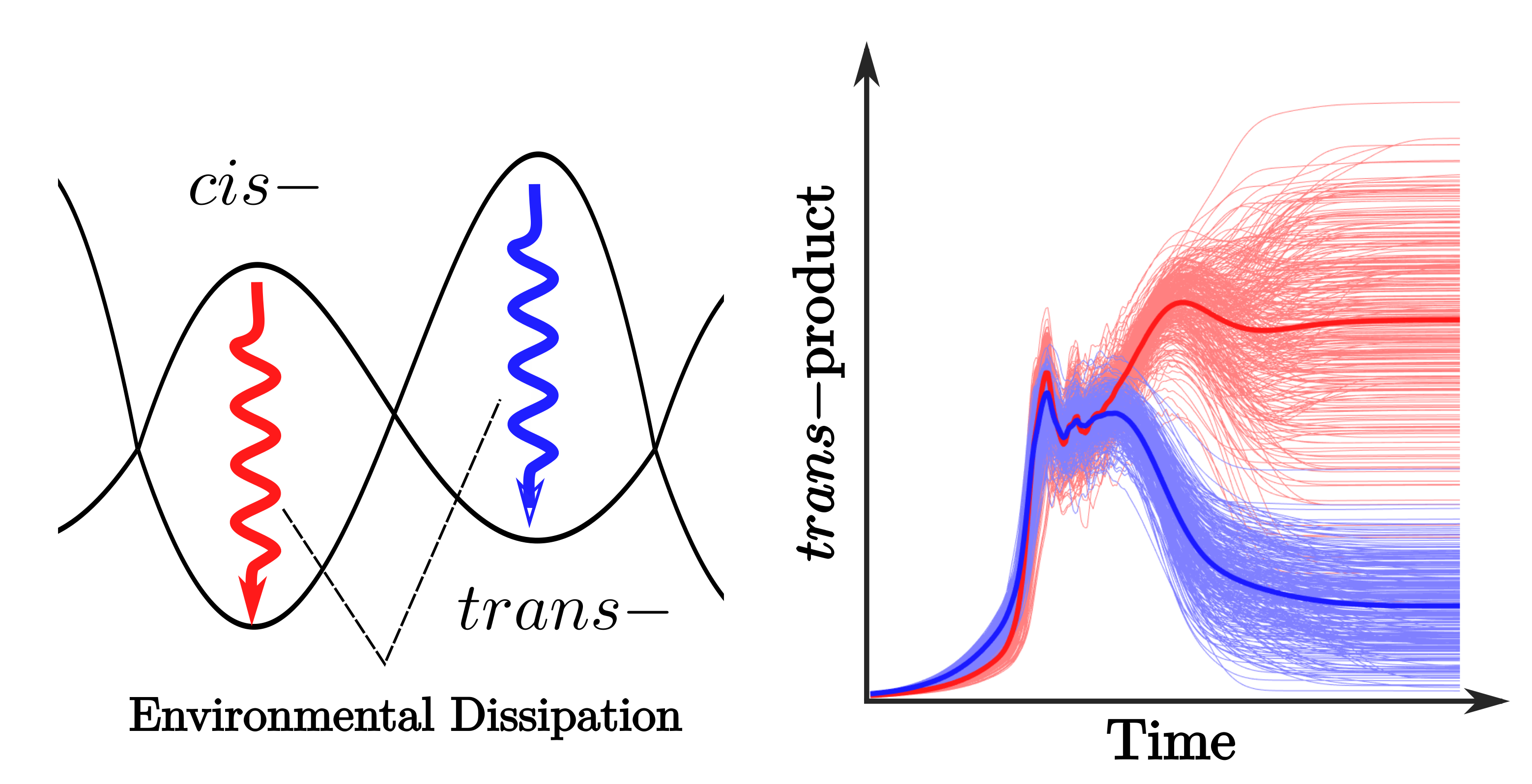}
  For Table of Contents Only
\end{tocentry}

\begin{abstract}
We simulate the nonequilibrium steady state \textit{cis-trans} photoisomerization of retinal chromophore in rhodopsin based on a two-state-two-mode model coupled to a thermal environment. By analyzing the systematic trends within an inhomogeneously broadened ensemble of systems, we find that the steady state reaction quantum yield (QY) correlates strongly with the excess energy above the crossing point of the system, in agreement with the prediction of the short time dynamical wavepacket picture. However, the nontrivial dependence of the QY on the system-environment interaction indicates that a pure dynamical picture is insufficient and that environment-induced partial internal energy redistribution takes place before the reaction concludes. These results imply that a proper treatment of the photoisomerization reaction, particularly its high QY, must account for the redistribution and dissipation of energy beyond the dynamical wavepacket motion that is typically employed in the literature and that is appropriate only in the transient regime.
\end{abstract}

\maketitle
\section{Introduction}
Much of the detailed information regarding biologically significant light induced processes (such as vision and photosynthesis) arises from modern laser experiments, where ultrafast pulses are used to activate the first steps in photosynthesis, vision, skin response, etc. These experiments initiate dynamics on ever decreasing time scales, with current studies on tens of femtoseconds or faster.\cite{Chergui2010ARPC,JACSChlorosome2012,schnedermann2015JACS,Duan17,Martin2017ChemRev,Oliver2018RSOS} The result is information on the nature of the system Hamiltonian and the system-environment interaction. However, natural light-induced processes, as we have repeatedly emphasized, take place under incoherent (e.g. sunlight) excitation where the molecules display totally different responses than that of  pulsed laser excitation.\cite{JiangBrumer1991JCP,Brumer2012,Han2013,BrumerPerspective2018,dodin2019,Dodin2021} Specifically, incoherent excitation leads to a nonequilibrium steady state (NESS), as distinct from time evolving coherent evolution.

We focus here on the first steps in vision, the \textit{cis}-\textit{trans} isomerization of retinal, a process that occurs via a conical intersection between the lowest two electronic states.\cite{Wald1968,DARTNALL1968,Hurley1977,kim2001} The analysis of ultrafast experimental results on such systems has yielded models and intuition based upon wave packet dynamics and upon an analogy to the Landau-Zener (LZ) model.\cite{schoenlein1991,wang1994,kim2003,schnedermann2015JACS,johnson2017} The LZ formula provides a theoretical foundation that relates transient dynamics, specifically the velocity of the wavepacket in the vicinity of the crossing point, to the transition probability at infinite time after crossing. However, utilizing this short time wavepacket picture to the interpretation of condensed phase photochemical reactions involving conical intersections is extremely challenging. The high dimensionality and strong dissipation afforded by the environment are expected to significantly alter this physical picture and, in the sense of evolutionary optimization, control the reaction dynamics.\cite{AoRammer1991PRB,KayanumaNakayama1998PRB,NalbachThorwart2010ChemPhys,Amro2014JCP,NovelliBelzigNitzan2015NJP} This includes the vibrational degrees of freedom (DOF) of the retinal residue other than the few key modes including the torsional, the hydrogen out-of-plane, and the bond length alternation modes,\cite{FaragJansenKnoester2018PCCP,Olivucci2S3M2019JPCA} as well as those belonging to the opsin protein pocket. More specifically, the rhodopsin protein pocket is known to critically affect both the static conformation of low-energy isomers and the dynamical processes relating them.\cite{Hurley1977,Ottolenghi1982,Sheves2013} These additional DOF introduce modifications to both the transient dynamics and, arguably more biologically relevant, the reaction quantum yield (QY) in the NESS. Consequently, establishing the connection between information retrieved in these two very different temporal regimes is crucial.

There has been a wealth of information on the photoisomerization of rhodopsin provided by short-time atomistic wavepacket simulations\cite{Ishida2012JPCB,Olivucci2017ChemRev,schnedermann2018} based on popular surface-hopping methods.\cite{ZhuNakamura1992JCP2,TullyPerspective2012JCP,Slavicek2020JCTC} These results are in general agreement with the corresponding transient nonlinear spectroscopic measurements. However, the connection between transient spectroscopic features, such as photoproduct rise time, and the steady state QY is less clear.\cite{schnedermann2018} Naively, a direct application of the LZ formula predicts higher transition probability (QY) with faster crossing speed: $P\propto e^{-A/v}$, where $A$ is a constant proportional to the coupling between the two states and $v$ is the rate at which the energy gap closes, which can be interpreted as the component of the wavepacket velocity along the gap-closing direction. It has been argued that the correlation of the results from the two temporal regimes can be recovered in part by accounting for the phase relationship between key local vibrational modes.\cite{ElTahawy2018JPCL} However, a few fundamental questions remain: First, it is unclear how the influence on QY from the temporal phase relationships among the various DOF established by coherent excitation survives in the natural conditions of excitations by incoherent light.\cite{Hoki2011,BrumerPerspective2018} Second, in adopting such a coherent wavepacket picture one assumes that the energy provided by the photoexcitation is conserved within these key modes before the reaction concludes. This is in sharp contrast to traditional rate theories where energy flow among various system DOF is assumed to be fast and efficient.\cite{GruebeleWolynes2004AccChemRes} Third, the vast Hilbert space for large molecules such as rhodopsin presents the possibility that the energy per mode is small while the overall energy is enough for the reaction to take place, rendering classical approximation invalid.\cite{XieDomcke2017JCP} 


Here we address the specific question of the extent to which wavepacket based intuition holds in the NESS that characterizes retinal isomerization in nature. We show that while the short-time wavepacket picture is capable of explaining some of our results in simulating rhodopsin photoisomerization under an open quantum system setup, the effect of energy exchange among various system DOF can be significant. As such, a dynamical, wavepacket-based physical picture is insufficient, and partial thermalization of internal energy distribution is critical in determining the systematic trends of steady state QY. 

This paper is organized as follows. We present the specific open quantum system model and the methodology in Sec.~\ref{sec:Method}. The main results and discussion are presented in Sec.~\ref{sec:ResultsDiscussion}. We conclude in Sec.~\ref{sec:Conclusion}.

\section{Model and Method}
\label{sec:Method}
\begin{figure}
    \centering
\includegraphics[width=14cm]{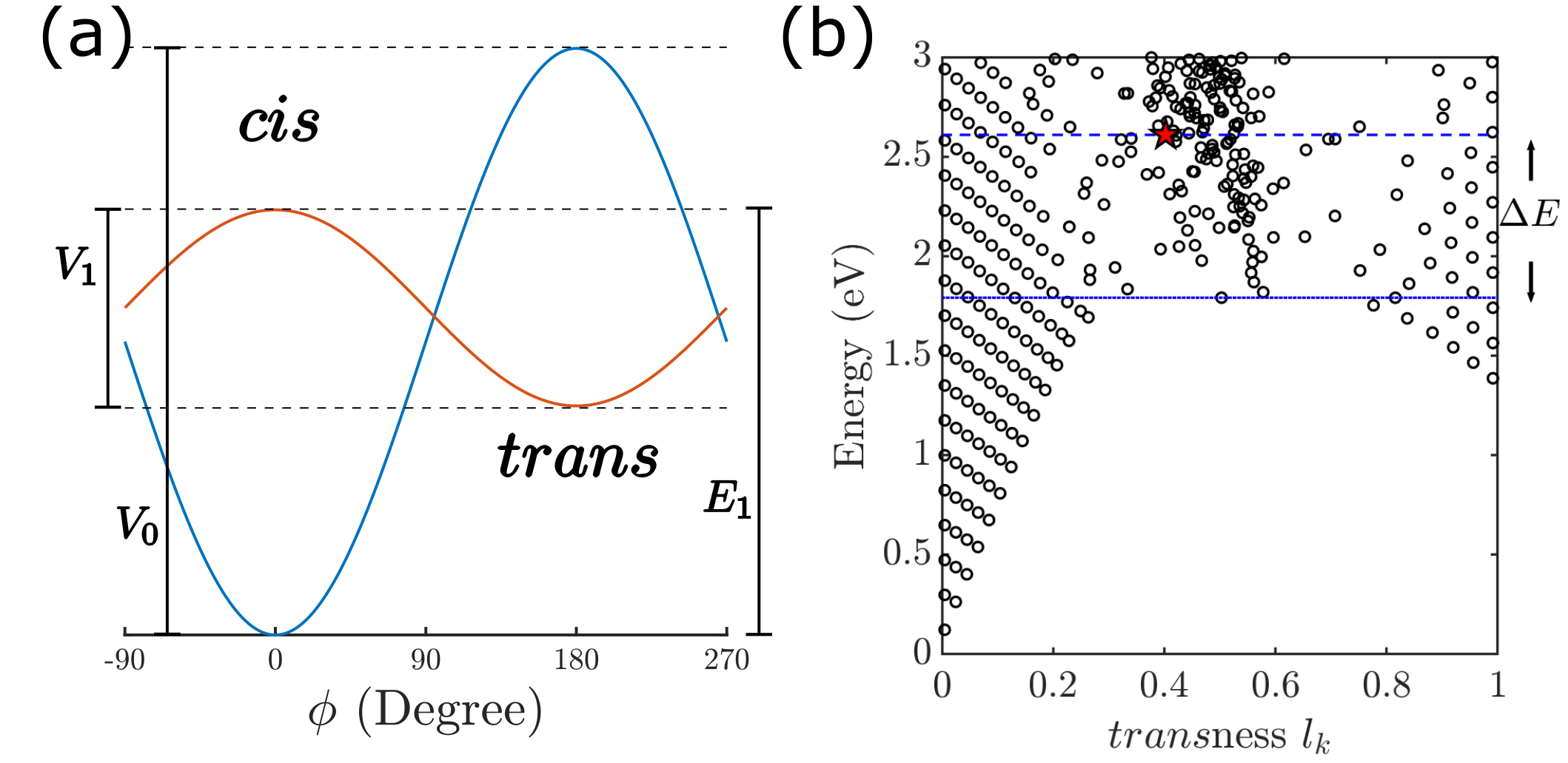}
    \caption{(a) Schematic drawing of the two diabatic surfaces as functions of the reaction coordinate $\phi$. (b) The energy eigenstate distribution $(l_k,~\epsilon_k)$ of the system Hamiltonian, Eq.~(\ref{eqn:Hs}). $\Delta E$ is the energy difference between the Franck-Condon state and the lowest energy eigenstate that is delocalized between the \textit{cis}- and \textit{trans}-wells (that is, with $l_k\approx0.5$).}
    \label{fig:F1}
\end{figure}

Consider first a general system-bath Hamiltonian that can be written as 
\begin{eqnarray}
H=H_\mathrm{s}+H_\mathrm{b}+H_\mathrm{sb}.
\end{eqnarray}
The system Hamiltonian $H_\mathrm{s}$ is composed of two diabatic electronic states $|0\rangle$ and $|1\rangle$, a reaction coordinate $\phi$, and a harmonic mode $x$:
\begin{eqnarray}
H_\mathrm{s}&=\sum_{n,n'=0,1}&\left[\left(\hat{T}+E_n+(-1)^n\frac{V_n}{2}(1-\cos\phi)+\frac{\omega x^2}{2}+\kappa x\delta_{n,1}\right)\delta_{n,n'}+\right.\nonumber\\
&&\left.\lambda x(1-\delta_{n,n'})\right]|n\rangle\langle n'|\label{eqn:Hs}
\end{eqnarray}
where $\hat{T}=-\frac{1}{2m}\frac{\partial^2}{\partial\phi^2}-\frac{\omega}{2}\frac{\partial^2}{\partial x^2}$ is the kinetic energy operator. All the parameters are taken from Refs.\cite{johnson2017} and \cite{balzer2005}, and are summarized in Table~\ref{tab:parameters}. This is referred to as a two-state-two-mode (2S2M) model, and a schematic drawing of the diabatic states as functions of $\phi$ is shown in Fig.\ref{fig:F1}(a). By expressing the system Hamiltonian in the Fourier grid basis\cite{MarstonBalintKurti1989JCP} in the $\phi$ coordinate (220 grid points) and in the number basis for the harmonic mode $x$ (truncated at $N_x=$20) and numerically diagonalizing it, we obtain its eigenvalues and eigenfunctions. As was discussed in our previous report,\cite{RetinalChaos1} a useful measure to understand the nature of the energy eigenfunctions of $H_\mathrm{s}$ is given by
\begin{eqnarray}
l_k&=&\frac{1}{2\pi}\int_0^{2\pi}d\phi~|\langle\phi|k\rangle|^2\cdot\frac{(1-\cos\phi)}{2}
\end{eqnarray}
where $|\phi\rangle$ is the position eigenfunction along the reaction coordinate $\phi$ and $|k\rangle$ is the $k^\mathrm{th}$ energy eigenfunction of $H_\mathrm{s}$ of energy $\epsilon_k$. This quantity measures the ``\textit{trans-}ness'' of the state in question: States completely localized in the \textit{cis}-well correspond to $l_k=0$, and those localized in the \textit{trans}-well have $l_k=1$. In Fig.\ref{fig:F1}(b) we plot all the energy eigenstates of $H_\mathrm{s}$ under 3 eV using $(l_k,\epsilon_k)$ as coordinates.  The brightest state (state with the largest amplitude composing the Franck-Condon state) is indicated by the red star, with its energy marked by the dashed line, and the solid line indicates the energy of the lowest energy state that has $|l_k-0.5|<0.1$. The latter can be interpreted as the ``point of no return'' in the relaxation dynamics, similar to the curve crossing point of the LZ model. The photoproduct population is defined as that projected onto the diabatic state $|1\rangle$ and within the range $\phi\in\left[\pi/2,3\pi/2\right)$. We focus the discussion on this quantity, as its steady state value defines the photoisomerization QY. 

\begin{table}[h]
\begin{tabular}{|c|c|c|c|c|c|c|}
\hline
Parameter\textsuperscript{\emph{a}}   (eV) & $E_0$       & $E_1$       & $V_0$             & $V_1$                    &  $m^{-1}$                         & $\omega$         \\ \hline
-                        & 0                & 2.58           & 3.56                &  1.19                      &  28.06$\cdot10^{-4}$       & 0.19           \\ \hline
-                        &$\kappa$    & $\lambda$ & $\eta_\phi$      & $\omega_{c,\phi}$ & $\eta_x$                          & $\omega_{c,x}$  \\ \hline
-                        & 0.19           &  0.19          &  0.15            &  0.071\textsuperscript{\emph{b}}                      &  0.1                                & 0.19                 \\ \hline
\end{tabular}
\caption{Table of parameters. }
\textsuperscript{\emph{a}}$\eta_\phi$ and $\eta_x$ are dimensionless and  all other parameters are in eV;\\
\textsuperscript{\emph{b}} When sampling the disorder in $V_0$ the value of $\omega_{c,\phi}=\sqrt{V_0/2m}$ is calculated accordingly.
\label{tab:parameters}
\end{table}

It has been estimated that the inhomogeneous contribution to the absorption linewidth of rhodopsin is on the order of 1000 cm$^{-1}$.\cite{hahn2002} In our simulations we found that the absorption linewidth can be accounted for by introducing a Gaussian disorder of magnitude $\sigma=2000$ cm$^{-1}$ to the parameter $E_1$, the optical gap of the \textit{cis}-conformer. The same disorder strength is also assumed for the \textit{trans}-optical gap $V_0-E_1+V_1$. For simplicity, when changing the $E_1$ value we simultaneously change $V_1$ the same amount in order to keep both the energy storage $(E_1-V_1)$ and the \textit{trans-}optical gap the same, while disorder in the \textit{trans-}optical gap is independently applied to $V_0$.  The effect of this change in $E_1$ and $V_0$ will be discussed later below. 


\subsection{System-Bath Coupling}
The system-bath coupling $H_\mathrm{sb}$ is approximated by the direct product form
\begin{eqnarray}
H_\mathrm{sb}&=&\sum_{\mathrm{b}} \hat{S}^\mathrm{(b)}\otimes\hat{B}^\mathrm{(b)},\label{eqn:FactorizedSB}
\end{eqnarray}
where $\hat{S}^\mathrm{(b)}$ and $\hat{B}^\mathrm{(b)}$ are operators that only depend on the system and the bath coordinates, respectively. The superscript $\mathrm{b}$ runs over all baths involved. The baths are described within the harmonic approximation, $H_\mathrm{ph}=\sum_k\omega_k~b^\dagger_kb_k$, and comprise two components:
\begin{eqnarray}
H_\mathrm{s-ph}&=&H_{\mathrm{s-}x}+H_{\mathrm{s}-\phi}\\
H_{\mathrm{s-}x}&=&S_x\cdot\sum_kg_{k,x}(b_{k,x}^\dagger+b_{k,x})\\
H_{\mathrm{s-}\phi}&=&S_\phi\cdot\sum_kg_{k,\phi}(b_{k,\phi}^\dagger+b_{k,\phi})
\label{eqn:sys-phi1}
\end{eqnarray}
where $g_{k,x}$ and $g_{k,\phi}$ are the displacement of the phonon mode $k$, $S_\phi=|1\rangle\langle1|\cdot(1-\cos\phi)$ and $S_x=|1\rangle\langle1|\cdot x$. We follow Stock et al.\cite{hahn2002,balzer2005} and adopt an Ohmic bath with exponential cut-off, $J_{y}(\omega)=\eta_y~\omega~e^{-\omega/\omega_{c,y}}$, where $y = x, \phi$, with parameters provided in Table~\ref{tab:parameters}. It is worth noting that the $\phi$-bath in Eq.~(\ref{eqn:sys-phi1}) couples primarily to the \textit{cis}-well owing to the $(1-\cos\phi)$ factor, whose significance will be discussed later.

To help understand the effects of system-phonon interaction, we also consider two additional forms for $H_{\mathrm{s-}\phi}$
\begin{eqnarray}
H_{\mathrm{s-}\phi'}&=&|1\rangle\langle1|\cdot\mathbb{I}_\phi\cdot\sum_kg_{k,\phi}(b_{k,\phi}^\dagger+b_{k,\phi})\label{eqn:sys-phi2}\\
H_{\mathrm{s-}\phi''}&=&|1\rangle\langle1|\cdot(1+\cos\phi)\cdot\sum_kg_{k,\phi}(b_{k,\phi}^\dagger+b_{k,\phi}).\label{eqn:sys-phi3}
\end{eqnarray}
Here the $\phi$ dependence in $H_{\mathrm{s-}\phi'}$ makes no distinction between the two wells, whereas $H_{\mathrm{s-}\phi''}$ couples primarily to the \textit{trans-}well.\cite{hahn2002} In essence, the form of $H_{\mathrm{s-}\phi}\propto S_{\phi}$ dictates how the phonon environment changes as a function of the reaction coordinate, as was carefully documented and discussed by Eyring et al.\cite{Eyring1980}. Here such changes are approximated by a position $(\phi,x)$-dependent operator applied to the same collection of environmental modes, specified by the spectral density.

The effect of system-bath interaction is treated with the standard Markovian Redfield quantum master equation formalism,\cite{Ishizaki2009} where the system reduced density matrix is subject to the following equation of motion in the system energy eigenbasis:
\begin{eqnarray}
\frac{\partial}{\partial t}\rho_{ij}&=&-i\omega_{ij}\rho_{ij}+\sum_\mathrm{b}\sum_{kl}\mathcal{R}_{ij,kl}^\mathrm{(b)}\rho_{kl},
\label{eqn:Redfield}
\end{eqnarray}
where $\omega_{ij}$ is the energy difference between the system eigenstates $i$ and $j$ and we have taken $\hbar=1$. Expressions for the relaxation tensor $\mathcal{R}_{ij,kl}$ can be found in, e.g., Ref.\cite{RetinalChaos1}. Assuming the initial state to be the Franck-Condon state from the overall ground state of $H_\mathrm{s}$ ($|g\rangle$), that is $|\mathrm{FC}\rangle=|1\rangle\langle0|g\rangle$, corresponding to coherent excitation in the impulsive limit, the system reduced density matrix is then propagated using Eq.~(\ref{eqn:Redfield}). Furthermore, we adopt the Bloch-secular approximation which decouples the population dynamics from the coherence and greatly simplifies the numerical effort required. This is justified since we are primarily interested in the long-time regime where this approximation introduces only minor deviations, as discussed in the literature.\cite{balzer2005,RetinalChaos1}.

\section{Results and Discussion}
\label{sec:ResultsDiscussion}
\subsection{Ensemble Average and the Short-Time Dynamical Wavepacket Perspective}
\begin{figure}
    \centering
\includegraphics[width=14cm]{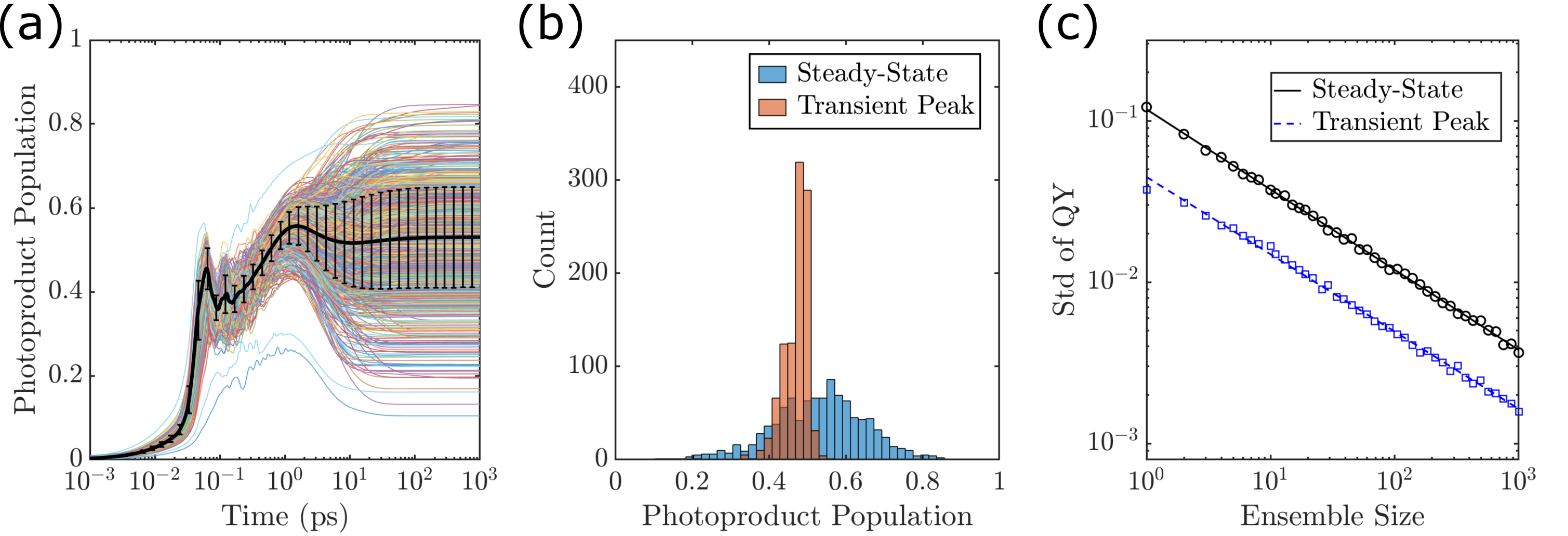}
    \caption{(a) 1000 time-dependent photoproduct population trajectories subject to a Gaussian-distributed $E_1$ parameter, representing the inhomogeneous broadening introduced by static disorder. The thick black line is the average with the error bar indicating the standard deviation. (b) The histograms of the steady state (recorded at 1 ns, shown in blue) and the transient peak photoproduct populations (brown), taken from the trajectories shown in Fig.~\ref{fig:F1}(b). The latter is defined by the magnitude of the first peak of the photoproduct population for each trajectory. (c) The standard deviations of the steady state (black circles) and the transient peak photoproduct populations as functions of ensemble size $N_\mathrm{E}$. The lines are $N_\mathrm{E}^{-0.5}$ fits to the data points.}
    \label{fig:F2}
\end{figure}

Natural retinal photoisomerization occurs in an ensemble with, \textit{e.g.}, a variable optical gap. It is within such an ensemble that the short-time dynamical wavepacket picture has been utilized. In Fig.\ref{fig:F2}(a) we show the dynamics initiated at the Franck-Condon states of an ensemble of a thousand 2S2M systems modeled by numerically solving Eq.~(\ref{eqn:Redfield}). Within this ensemble we sample a random optical gap parameter ($E_1$) drawn from a Gaussian distribution of width 2000 cm$^{-1}$, corresponding to 10\% of the mean optical gap. The time-dependent mean and standard deviation of the dynamics are shown in thick black line. The averaged dynamics shows a steep rise of photoproduct population on the time scale of $<100$ fs, followed by oscillatory features up to $\sim300$ fs. Both of these features have been well documented in pulsed laser experiments, reflecting the coherent vibronic wavepacket motion as initiated therein.\cite{schoenlein1991,wang1994,prokhorenko2006,schnedermann2015JACS,schnedermann2018} On the other hand, the mean photoproduct population continues to change after 1 ps and stabilizes to the steady state value $\sim 0.55\%$ at around 100 ps for the current model, while its standard deviation grows significantly, up to $\sim10\%$ at steady state. As we recently reported,\cite{RetinalChaos1} this phenomenon indicates the extreme sensitivity of nonadiabatic dynamics to system parameters for a large family of theoretical models commonly employed to simulate various photochemical systems in gas phase as well as in condensed phase.\cite{KoppelDomckeCederbaum1984ACP,Heller1990,LeitnerCederbaum1996,DomckeZhao2016}

In most retinal experimental setups, only the mean value of the quantum yield is reported. By contrast, here we show the standard deviation of the photoproduct population in Fig.~\ref{fig:F2}(a) and those of its short-time peak values and steady state counterparts in Fig.~\ref{fig:F2}(b), giving far greater insight into the ensemble dependence. The effect of ensemble averaging is seen more directly in Fig.~\ref{fig:F2}(c), which shows the dependence of the standard deviations of the steady state and the transient peak photoproduct populations on the number of independent trajectories that are averaged over. The values in Fig.\ref{fig:F2}(b) are understood as single-molecule trajectories in which systems with different parameters can be resolved individually. As such, they correspond to the left most data point in Fig. 2(c), where $N_\mathrm{E}=1$. The data points at, for example, $N_\mathrm{E}=10$ are calculated by first averaging individual trajectories in groups of ten and then taking the statistical measures with respect to the averaged trajectories. \footnote{The transient peak values of the standard deviation in Fig.\ref{fig:F2}(b) are evaluated by averaging over the photoproduct population at the first peaks of individual trajectories. This is different from that shown in Fig.\ref{fig:F1}(c) where the standard deviation is taken at the same time stamp across all trajectories and, therefore, does not account for the out-of-sync effect among them.}  In essence, parametric sensitivity is attenuated upon averaging over a macroscopic number of molecules. (This is especially true for biological systems with inhomogeneity from slow and large-scale conformation changes inherent to their protein environment, as in rhodopsin.) As noted above, it is in the ensemble-averaged domain that the short-time dynamical wavepacket approaches have been advocated. They are not expected to \textit{a priori} predict the extreme variation of QY to parameters evident in Fig.\ref{fig:F2}(a). 

It can be seen that both the transient peak and the steady state values follow an $N_\mathrm{E}^{-0.5}$ scaling, reflecting the Poisson statistics typically found in single-molecule/small ensemble measurements where individual events are uncorrelated.\cite{CaoBawendi2007PRB} Also, the transient peak standard deviation is consistently smaller than the steady state counterpart by a factor of $2.7$, \textit{i.e.} the parametric sensitivity is much more significant at the steady state. The deviations become undetectable ($<1\%$) for an ensemble size beyond $\sim$100 for the current model, a lower bound for the ensemble size that the short-time dynamical wavepacket perspective can be explored experimentally. 

Here, however, we are primarily interested in the systematic features of steady state photoisomerization reaction QY subject to ensemble averaging, which concerns the majority of biological circumstances. In particular, our emphasis is on the effects of both the general structure of the system's vibronic topography and the influence of system-bath interaction, discussed in the following.

\subsection{System Vibronic Landscape}
The short-time dynamical wavepacket perspective has been adopted to explain the excitation wavelength dependence of the photoisomerization QY of rhodopsin, where the QY was found to decrease upon redshifting the excitation wavelength from the absorption peak. \cite{kim2001,kim2003} Does this particular feature extend to the steady state? 

\begin{figure}
    \centering
\includegraphics[width=14cm]{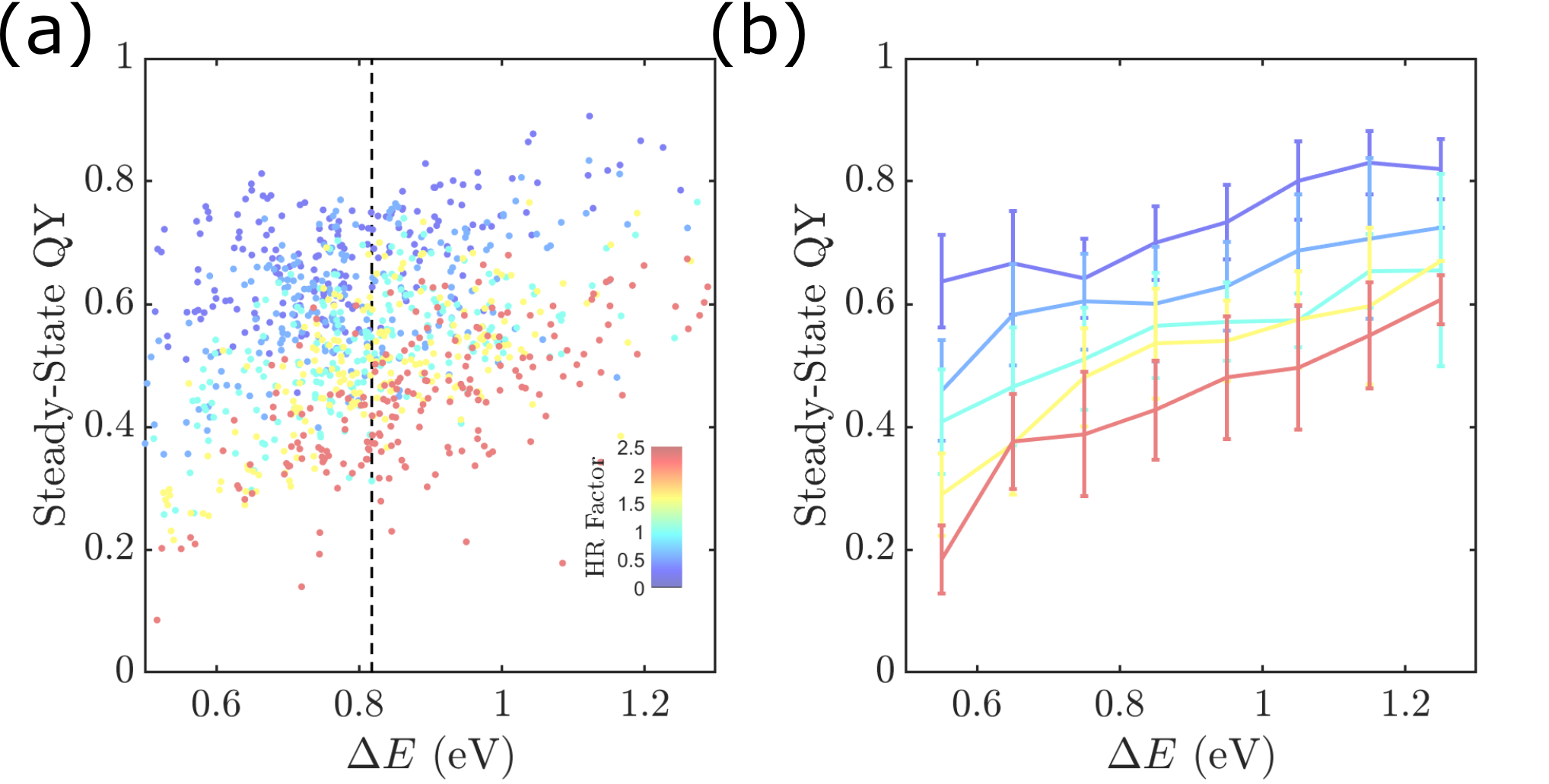}
    \caption{(a) Scatter plot of the steady state QY with respect to $\Delta E$ for a collection of 2S2M models with inhomogeneously broadened optical gaps and varying $\kappa$ parameter while holding all other parameters constant. The data points are grouped and color coded according to their $\kappa$ values. The dashed line indicates $\Delta E^{(0)}=0.82$ eV, the value for the original model parameter set used in Ref.\cite{johnson2017}. (b) The data in (a) window averaged in steps of $0.1$ eV in $\Delta E$, each group contains up to 50 data points. The error bars show the extent of one standard deviation.}
    \label{fig:F3}
\end{figure}

In simulating the steady state QY of the 2S2M model, we associate a higher crossing velocity to a larger excess energy $\Delta E$, defined in Fig.~\ref{fig:F1}(b), the energy difference between the Franck-Condon state and the lowest energy state that delocalizes between the reactant and the product wells. The latter can be seen as the ``point of no return'' of the photoisomerization dynamics provided that the thermally activated contribution is negligible.\footnote{$\Delta E$ is similar to the excess energy in Ref.\cite{kim2003}. However therein both the ground and the excited states are treated as harmonic multi-dimensional surfaces.} Intuitively, a larger $\Delta E$ corresponds to a larger crossing velocity in the LZ model.\footnote{We note that this analogy is not exact as the system dynamics is studied with a quantum master equation, that the system is generally described by a mixed state. The notion of wavepacket velocity is lost when the inter-state coherence is destroyed by the couplings to the thermal environment. } Fig.~\ref{fig:F3}(a) examines the relationship between $\Delta E$ and the steady state reaction QY. Specifically, we plot the steady state QY against $\Delta E$ for a collection of 2S2M systems with Gaussian distributed optical gap, as described in the previous section, keeping all other parameters the same but one ($\kappa$, the relative displacement of the $x$ mode). In particular, the data is divided into subgroups (color coded) according to the magnitude of the Huang-Rhys factor $(\kappa/\omega)^2$. We also present the window-averaged data to the subgroups with increasing $\Delta E$ value in Fig.~\ref{fig:F3}(b).

While the data points are scattered extensively, manifesting parametric sensitivity,\cite{RetinalChaos1} we find a generic proportionality between steady state QY and $\Delta E$, supporting the short-time dynamical wavepacket perspective. In addition, with increasing $\kappa$ there is a trend towards decreasing QY. This is also in accordance with the dynamical wavepacket picture, where the excess energy deposited into the second mode would be less conducive to the reaction QY. Experimentally, parameters such as $\Delta E$ and $\kappa$ can be accessed by analyzing the vibrational spectra and by estimating the corresponding Franck-Condon factors.\cite{kim2003}

Would this perspective extend to larger systems? We note that there have been recent accounts of a different vibronic effects by introducing and analysing the dynamics along a third mode in the photoisomerization process.\cite{schnedermann2018,ElTahawy2018JPCL} Specifically, it was found that the \textit{short-time} QY is highly correlated with the relative phases of wavepacket motion along the three key local coordinates of the photoisomerization reaction: The reactive carbon and hydrogen torsional angles and the bond length alternation mode. While characterizing such dynamical effect on a higher dimensional model such as a two-state-three-mode model\cite{Olivucci2S3M2019JPCA} is an ongoing effort of the authors, we expect a trend similar to that shown in Fig.~\ref{fig:F3}. In this regard, a few other points are worth considering. First, while it is clear that wavepacket motion initiated at the Franck-Condon state remains coherent in the transient regime, and hence would be sensitive to the third mode, the reaction dynamics typically reaches the steady state limit in the 10 ps regime, where vibronic coherence is entirely lost. Hence considerations regarding the \textit{dynamics} of 3-mode case would likely be washed out. Second, as can be seen in Fig.~\ref{fig:F2}(a) and extensively discussed previously,\cite{RetinalChaos1} the steady state QY is only weakly correlated with short-time dynamics. Lastly, an LZ analysis is indeed less useful when comparing different systems such as those studied in Ref.\cite{schnedermann2018} (deuteration of retinal chromophore) and in Ref.\cite{ElTahawy2018JPCL} (different animal rhodopsins), where changes made to the systems are beyond perturbative. In such cases the changes in the vibronic topography is large so that the exponential factor $A$ in the LZ expression $P\propto e^{-A/v}$ can change significantly, and the correlation between the crossing speed $v$ and QY can be obscured. On the other hand, an LZ analysis of QY \textit{within} an inhomogeneously broadened ensemble of the same system would reveal the connection between $v$ and QY more directly, as is done here and in Ref.\cite{kim2003}.

It should be mentioned that there have been numerous efforts in augmenting the original 1D LZ theory to account for higher dimensional systems and/or dissipative environment, both analytically\cite{AoRammer1991PRB,KayanumaNakayama1998PRB,Amro2014JCP} and numerically.\cite{NalbachThorwart2010ChemPhys,LandrySubotnik2011JCP,Ashhab2014PRA,NovelliBelzigNitzan2015NJP,Slavicek2020JCTC} In various situations, particularly when the additional DOF or the environment assert significant influence compared to the nonadiabatic coupling between the two states, it is known that the original LZ formula can be unreliable, such as the prediction of a non-monotonic dependence on the crossing speed.\cite{Amro2014JCP,NovelliBelzigNitzan2015NJP} It would be informative to apply or generalize these results and methods to the study of rhodopsin photoisomerization provided that the incoherent excitation as well as the steady state condition are accounted for.

\subsection{System-Environment Coupling}
\begin{figure}
    \centering
\includegraphics[width=14cm]{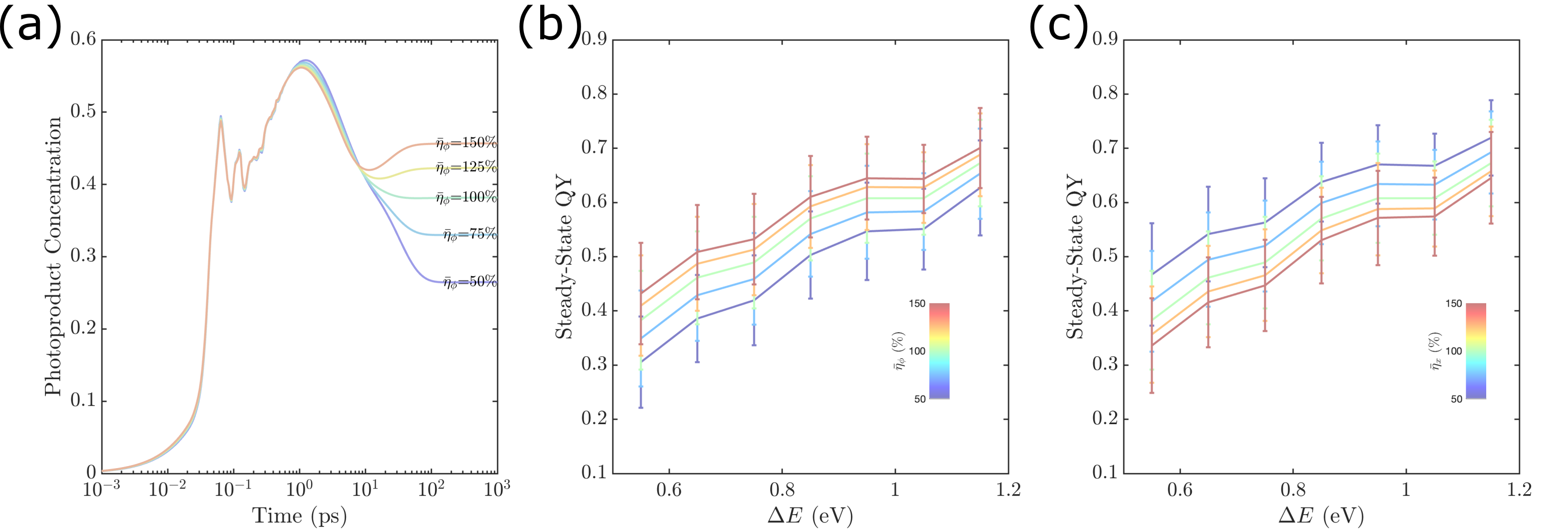}
    \caption{(a) Time-dependent photoproduct population of a sample system subject to different system-environment coupling magnitudes $\bar{\eta}_\phi$ while keeping all other parameters the same. The bottom (top) trajectory is $\bar{\eta}_\phi=50\%$  ($\bar{\eta}_\phi=150\%$), with $25\%$ increment for each successive trajectory. (b) Window averaged data as functions of the relative coupling strength in $H_{\mathrm{s}-\phi}$. The dependence on the coupling strength prefactor $\bar{\eta}_\phi$ is color coded. (c) Window averaged data as functions of the relative coupling strength in $H_{\mathrm{s}-x}$.  The dependence on the coupling strength prefactor $\bar{\eta}_x$ is color coded.}
    \label{fig:F4}
\end{figure}

The above analysis, while consistent with the short time wavepacket perspective, is not the full physical picture required to understand our results. Fig.~\ref{fig:F4} shows the dependence of the photoproduct population on the system-environment dissipation strength. Here we vary the prefactor $\bar{\eta}_y$ of the system-environment coupling term $H_{\mathrm{s-}y}$ in the form of $J_y(\omega)=\bar{\eta}_y\cdot\left[\eta_y~\omega~e^{-\omega/\omega_{c,y}}\right]$, where $y=\phi, x$. In Fig.~\ref{fig:F4}(a) we show a set of sample trajectories calculated using the same system parameters with varying $\bar{\eta}_\phi$ while keeping all other parameters the same. It is clear that while the transient dynamics is nearly unaffected (up to 10 ps), the long-time dynamics is sensitive to $\bar{\eta}_\phi$, a sensitivity that is completely missing in the short-time wavepacket dynamics picture.

For a more systematic study, in Fig.\ref{fig:F4}(b) we show the window averaged data (as introduced in Fig.~\ref{fig:F3}(b)) as functions $\bar{\eta}_\phi$, as well as the dependence on $\bar{\eta}_x$ in Fig.\ref{fig:F4}(c). The results show that while each set of data the reaction QY is generally proportional to $\Delta E$ consistent with the dynamical wavepacket picture, there is also a systematic increase of QY with increasing $\bar{\eta}_\phi$ and a decrease of QY with increasing $\bar{\eta}_x$. This suggests that the energy exchange between the two modes during the relaxation process, redistributing and dissipating the excess energy $\Delta E$, is significant. Specifically, our results imply that with increasing $\bar{\eta}_\phi$ a larger part of the excess energy is converted into the reaction coordinate, facilitating the conversion to the photoproduct, whereas the opposite is true for increasing $\bar{\eta}_x$. By contrast, a dynamical wavepacket picture would expect a negligible exchange of energy among reactive and unreactive modes. That is, a wavepacket picture cannot explain these results; \textit{i.e.} partial thermalization among the modes occurs during the photoisomerization process. 

\begin{figure}
    \centering
\includegraphics[width=14cm]{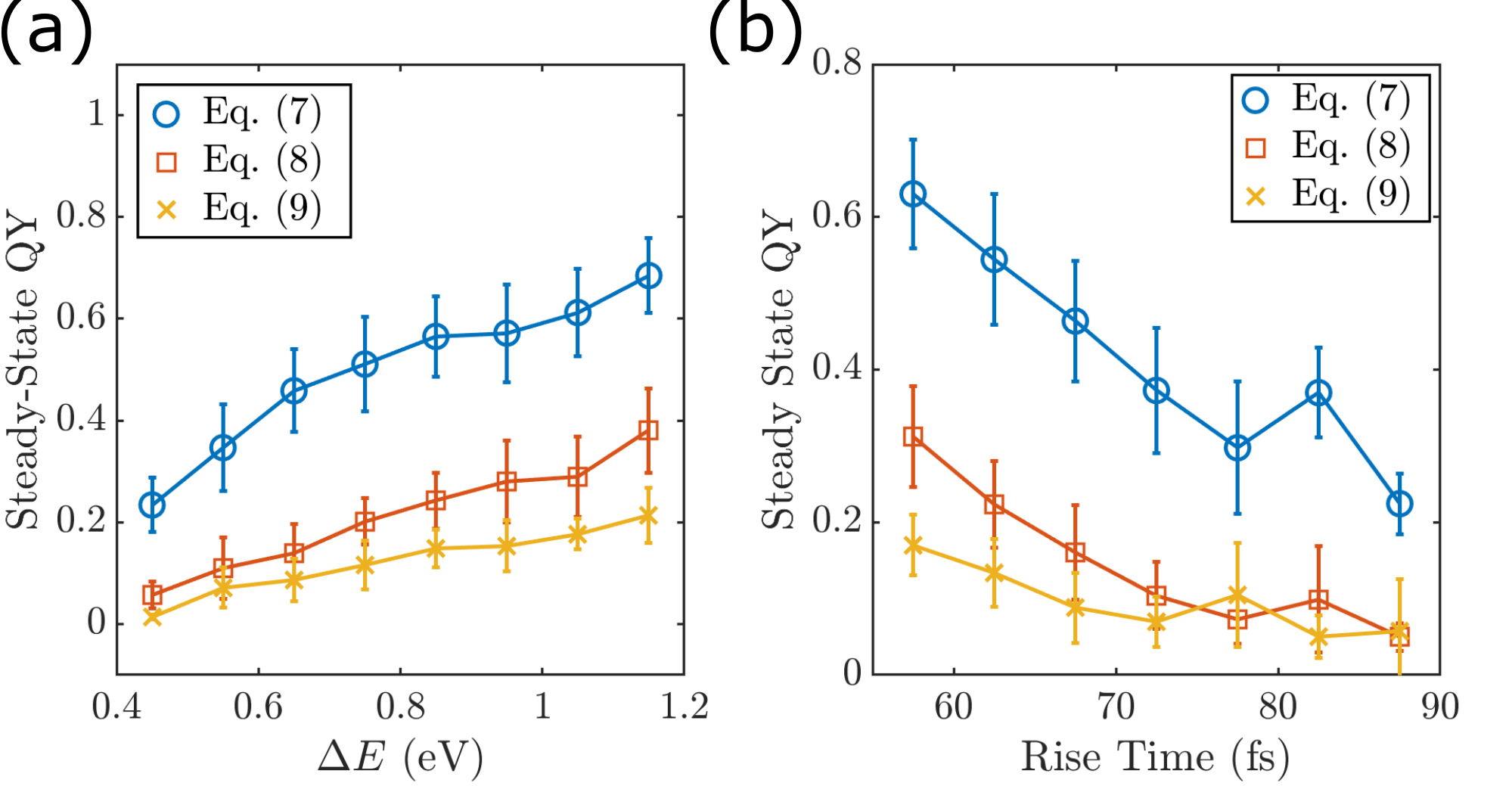}
    \caption{(a) Window average data as in Fig.~\ref{fig:F3}(b) where the system part of the system-environment coupling $S_\phi$ is changed as in Eqs.~(\ref{eqn:sys-phi1})-(\ref{eqn:sys-phi3}). Specifically, Eq.~(\ref{eqn:sys-phi1}) corresponds to stronger damping in the \textit{trans-}region, Eq.~(\ref{eqn:sys-phi3}) corresponds to stronger damping in the \textit{cis-}region, and Eq.~(\ref{eqn:sys-phi2}) is unbiased damping. (b) The same set of data in part (a) plotted against the photoproduct rise time instead. The rise time is defined in conjunction to the transient peak photoproduct population shown in Fig.\ref{fig:F2}, which is the time when the first peak in the photoproduct population is located. }
    \label{fig:F5}
\end{figure}

The effect of energy exchange also manifests in the functional form of the system-environment coupling, especially its dependence on the reaction coordinate as in Eqs.~(\ref{eqn:sys-phi1}) - (\ref{eqn:sys-phi3}). Some computational evidence is shown in Fig.~\ref{fig:F5}(a). In all three cases the steady state QY is found to be roughly proportional to the excess energy $\Delta E$, consistent with the dynamical wavepacket picture. However, the effect of partial thermalization is evident insofar as the steady state QY is significantly altered upon changing $S_\phi$.

Such partial thermalization is, as expected, most consequential beyond the transient regime. In Fig.~\ref{fig:F5}(b) we compare the photoproduct rise time among the three different system-environment coupling models. While the rise time indeed correlates inversely with the steady state QY, again in agreement with the dynamical wavepacket picture, the strong correlation between the form of the system-environment coupling and the steady state QY indicates a deficiency in the simple wavepacket picture. In fact, using a simpler two-state-one-mode model for rhodopsin, and considering the reverse reaction, Hahn and Stock showed that nearly all initial state dependence is erased at the NESS. \cite{hahn2002} 


\section{Conclusion}
\label{sec:Conclusion}
Pulsed laser studies of light-induced biological processes provide deep insights into the nature of the Hamiltonian and its coupling to the environment, through transient dynamics induced by the pulsed excitation. Interpretations of these experiments, based on a short-time dynamical wavepacket picture lend insight into the physics of the ongoing process. Here, we focused on the photoisomerization of rhodopsin, where interpretations of experiment have been closely linked to this dynamical perspective. In nature, however, this isomerization occurs in a non-equilibrium  steady state (NESS) that is  induced by incident incoherent (e.g. solar) radiation and the phonon environment, as opposed to pulsed coherent light. The resultant molecular behavior is dramatically different. The key issue addressed here is the extent to which intuition gained from the wavepacket picture carries over to an understanding of the quantum yield (the primary quantity of interest) in the NESS. Features such as the positive correlation between the quantum yield, the excess energy $\Delta E$, and inverse of the photoproduct rise time do persist in the steady state. However, features such as the relative strength and the functional form of the system-environment couplings, among other qualitative and quantitative features of the system-environment interaction, are distinctly different in the NESS than in the short time pulsed laser domain. In essence, interactions of the system with the environment, which can strongly affect the system QY, are manifest in the NESS and not in the transient dynamics. These conclusions are significant and are expected to carry over to other natural processes  where the light-induced NESS is operative.

In addition to these results we note that current experiments, which naturally entail an ensemble average over system properties, do not observe the extensive parametric sensitivity that we reported upon earlier, i.e. that the retinal QY depends strongly on system parameters. Here we have shown that the parametric sensitivity vanishes as $N_\mathrm{E}^{1/2}$, where $N_\mathrm{E}$ is the number of molecules in the ensemble. In addition to explaining why parametric sensitivity has not been observed, these results provide conditions under which experiments over reduced sized ensembles are expected to demonstrate this sensitivity. 

\section*{Acknowledgment} 
This work was supported by the U.S. Air Force Office of Scientific Research (AFOSR) under grant number FA9550-20-1-0354.

\bibliography{LH1RC}

\newcommand{\noopsort}[1]{} \newcommand{\printfirst}[2]{#1}
  \newcommand{\singleletter}[1]{#1} \newcommand{\switchargs}[2]{#2#1}
\providecommand{\latin}[1]{#1}
\providecommand*\mcitethebibliography{\thebibliography}
\csname @ifundefined\endcsname{endmcitethebibliography}
  {\let\endmcitethebibliography\endthebibliography}{}
\begin{mcitethebibliography}{54}
\providecommand*\natexlab[1]{#1}
\providecommand*\mciteSetBstSublistMode[1]{}
\providecommand*\mciteSetBstMaxWidthForm[2]{}
\providecommand*\mciteBstWouldAddEndPuncttrue
  {\def\EndOfBibitem{\unskip.}}
\providecommand*\mciteBstWouldAddEndPunctfalse
  {\let\EndOfBibitem\relax}
\providecommand*\mciteSetBstMidEndSepPunct[3]{}
\providecommand*\mciteSetBstSublistLabelBeginEnd[3]{}
\providecommand*\EndOfBibitem{}
\mciteSetBstSublistMode{f}
\mciteSetBstMaxWidthForm{subitem}{(\alph{mcitesubitemcount})}
\mciteSetBstSublistLabelBeginEnd
  {\mcitemaxwidthsubitemform\space}
  {\relax}
  {\relax}

\bibitem[Bressler and Chergui(2010)Bressler, and Chergui]{Chergui2010ARPC}
Bressler,~C.; Chergui,~M. Molecular Structural Dynamics Probed by Ultrafast
  X-Ray Absorption Spectroscopy. \emph{Ann. Rev. Phys. Chem.} \textbf{2010},
  \emph{61}, 263--282\relax
\mciteBstWouldAddEndPuncttrue
\mciteSetBstMidEndSepPunct{\mcitedefaultmidpunct}
{\mcitedefaultendpunct}{\mcitedefaultseppunct}\relax
\EndOfBibitem
\bibitem[Dostal \latin{et~al.}({2012})Dostal, Mancal, Augulis, Vacha, Psencik,
  and Zigmantas]{JACSChlorosome2012}
Dostal,~J.; Mancal,~T.; Augulis,~R.; Vacha,~F.; Psencik,~J.; Zigmantas,~D.
  {Two-Dimensional Electronic Spectroscopy Reveals Ultrafast Energy Diffusion
  in Chlorosomes}. \emph{{J. Amer. Chem. Soc.}} \textbf{{2012}}, \emph{{134}},
  {11611}\relax
\mciteBstWouldAddEndPuncttrue
\mciteSetBstMidEndSepPunct{\mcitedefaultmidpunct}
{\mcitedefaultendpunct}{\mcitedefaultseppunct}\relax
\EndOfBibitem
\bibitem[Schnedermann \latin{et~al.}(2015)Schnedermann, Liebel, and
  Kukura]{schnedermann2015JACS}
Schnedermann,~C.; Liebel,~M.; Kukura,~P. Mode-Specificity of Vibrationally
  Coherent Internal Conversion in Rhodopsin during the Primary Visual Event.
  \emph{J. Amer. Chem. Soc.} \textbf{2015}, \emph{137}, 2886--2891\relax
\mciteBstWouldAddEndPuncttrue
\mciteSetBstMidEndSepPunct{\mcitedefaultmidpunct}
{\mcitedefaultendpunct}{\mcitedefaultseppunct}\relax
\EndOfBibitem
\bibitem[Duan \latin{et~al.}(2017)Duan, Prokhorenko, Cogdell, Ashraf, Stevens,
  Thorwart, and Miller]{Duan17}
Duan,~H.-G.; Prokhorenko,~V.~I.; Cogdell,~R.~J.; Ashraf,~K.; Stevens,~A.~L.;
  Thorwart,~M.; Miller,~R. J.~D. Nature does not rely on long-lived electronic
  quantum coherence for photosynthetic energy transfer. \emph{Proc. Natl. Acad.
  Sci.} \textbf{2017}, \emph{114}, 8493--8498\relax
\mciteBstWouldAddEndPuncttrue
\mciteSetBstMidEndSepPunct{\mcitedefaultmidpunct}
{\mcitedefaultendpunct}{\mcitedefaultseppunct}\relax
\EndOfBibitem
\bibitem[Nisoli \latin{et~al.}({2017})Nisoli, Decleva, Calegari, Palacios, and
  Martin]{Martin2017ChemRev}
Nisoli,~M.; Decleva,~P.; Calegari,~F.; Palacios,~A.; Martin,~F. {Attosecond
  Electron Dynamics in Molecules}. \emph{{Chem. Rev.}} \textbf{{2017}},
  \emph{{117}}, {10760}\relax
\mciteBstWouldAddEndPuncttrue
\mciteSetBstMidEndSepPunct{\mcitedefaultmidpunct}
{\mcitedefaultendpunct}{\mcitedefaultseppunct}\relax
\EndOfBibitem
\bibitem[Oliver({2018})]{Oliver2018RSOS}
Oliver,~T. A.~A. {Recent advances in multidimensional ultrafast spectroscopy}.
  \emph{{R. Soc. Open Sci.}} \textbf{{2018}}, \emph{{5}}\relax
\mciteBstWouldAddEndPuncttrue
\mciteSetBstMidEndSepPunct{\mcitedefaultmidpunct}
{\mcitedefaultendpunct}{\mcitedefaultseppunct}\relax
\EndOfBibitem
\bibitem[Jiang and Brumer(1991)Jiang, and Brumer]{JiangBrumer1991JCP}
Jiang,~X.; Brumer,~P. Creation and dynamics of molecular states prepared with
  coherent vs partially coherent pulsed light. \emph{J. Chem. Phys.}
  \textbf{1991}, \emph{94}, 5833--5843\relax
\mciteBstWouldAddEndPuncttrue
\mciteSetBstMidEndSepPunct{\mcitedefaultmidpunct}
{\mcitedefaultendpunct}{\mcitedefaultseppunct}\relax
\EndOfBibitem
\bibitem[Brumer and Shapiro(2012)Brumer, and Shapiro]{Brumer2012}
Brumer,~P.; Shapiro,~M. Molecular response in one-photon absorption via natural
  thermal light vs. pulsed laser excitation. \emph{Proc. Nat. Acad. Sci.}
  \textbf{2012}, \emph{109}, 19575--19578\relax
\mciteBstWouldAddEndPuncttrue
\mciteSetBstMidEndSepPunct{\mcitedefaultmidpunct}
{\mcitedefaultendpunct}{\mcitedefaultseppunct}\relax
\EndOfBibitem
\bibitem[Han \latin{et~al.}(2013)Han, Shapiro, and Brumer]{Han2013}
Han,~A.~C.; Shapiro,~M.; Brumer,~P. Nature of Quantum States Created by One
  Photon Absorption: Pulsed Coherent vs Pulsed Incoherent Light. \emph{J. Phys.
  Chem. A} \textbf{2013}, \emph{117}, 8199--8204\relax
\mciteBstWouldAddEndPuncttrue
\mciteSetBstMidEndSepPunct{\mcitedefaultmidpunct}
{\mcitedefaultendpunct}{\mcitedefaultseppunct}\relax
\EndOfBibitem
\bibitem[Brumer(2018)]{BrumerPerspective2018}
Brumer,~P. Shedding (Incoherent) Light on Quantum Effects in Light-Induced
  Biological Processes. \emph{J. Phys. Chem. Lett.} \textbf{2018}, \emph{9},
  2946--2955\relax
\mciteBstWouldAddEndPuncttrue
\mciteSetBstMidEndSepPunct{\mcitedefaultmidpunct}
{\mcitedefaultendpunct}{\mcitedefaultseppunct}\relax
\EndOfBibitem
\bibitem[Dodin and Brumer(2019)Dodin, and Brumer]{dodin2019}
Dodin,~A.; Brumer,~P. Light-induced processes in nature: Coherences in the
  establishment of the nonequilibrium steady state in model retinal
  isomerization. \emph{J. Chem. Phys.} \textbf{2019}, \emph{150}, 184304\relax
\mciteBstWouldAddEndPuncttrue
\mciteSetBstMidEndSepPunct{\mcitedefaultmidpunct}
{\mcitedefaultendpunct}{\mcitedefaultseppunct}\relax
\EndOfBibitem
\bibitem[Dodin and Brumer({2022})Dodin, and Brumer]{Dodin2021}
Dodin,~A.; Brumer,~P. Noise-induced coherence in molecular processes. \emph{J.
  Phys. B} \textbf{{2022}}, \emph{{54}}, {223001}\relax
\mciteBstWouldAddEndPuncttrue
\mciteSetBstMidEndSepPunct{\mcitedefaultmidpunct}
{\mcitedefaultendpunct}{\mcitedefaultseppunct}\relax
\EndOfBibitem
\bibitem[Wald(1968)]{Wald1968}
Wald,~G. The Molecular Basis of Visual Excitation. \emph{Nature} \textbf{1968},
  \emph{219}, 800\relax
\mciteBstWouldAddEndPuncttrue
\mciteSetBstMidEndSepPunct{\mcitedefaultmidpunct}
{\mcitedefaultendpunct}{\mcitedefaultseppunct}\relax
\EndOfBibitem
\bibitem[Dartnall(1968)]{DARTNALL1968}
Dartnall,~H. The photosensitivities of visual pigments in the presence of
  hydroxylamine. \emph{Vision research} \textbf{1968}, \emph{8}, 339--358\relax
\mciteBstWouldAddEndPuncttrue
\mciteSetBstMidEndSepPunct{\mcitedefaultmidpunct}
{\mcitedefaultendpunct}{\mcitedefaultseppunct}\relax
\EndOfBibitem
\bibitem[Hurley \latin{et~al.}(1977)Hurley, Ebrey, Honig, and
  Ottolenghi]{Hurley1977}
Hurley,~J.~B.; Ebrey,~T.~G.; Honig,~B.; Ottolenghi,~M. Temperature and
  wavelength effects on the photochemistry of rhodopsin, isorhodopsin,
  bacteriorhodopsin and their photoproducts. \emph{Nature} \textbf{1977},
  \emph{270}, 540--542\relax
\mciteBstWouldAddEndPuncttrue
\mciteSetBstMidEndSepPunct{\mcitedefaultmidpunct}
{\mcitedefaultendpunct}{\mcitedefaultseppunct}\relax
\EndOfBibitem
\bibitem[Kim \latin{et~al.}(2001)Kim, Tauber, and Mathies]{kim2001}
Kim,~J.~E.; Tauber,~M.~J.; Mathies,~R.~A. Wavelength dependent cis-trans
  isomerization in vision. \emph{Biochem.} \textbf{2001}, \emph{40},
  13774--13778\relax
\mciteBstWouldAddEndPuncttrue
\mciteSetBstMidEndSepPunct{\mcitedefaultmidpunct}
{\mcitedefaultendpunct}{\mcitedefaultseppunct}\relax
\EndOfBibitem
\bibitem[Schoenlein \latin{et~al.}(1991)Schoenlein, Peteanu, Mathies, and
  Shank]{schoenlein1991}
Schoenlein,~R.; Peteanu,~L.; Mathies,~R.; Shank,~C. The first step in vision:
  femtosecond isomerization of rhodopsin. \emph{Science} \textbf{1991},
  \emph{254}, 412--415\relax
\mciteBstWouldAddEndPuncttrue
\mciteSetBstMidEndSepPunct{\mcitedefaultmidpunct}
{\mcitedefaultendpunct}{\mcitedefaultseppunct}\relax
\EndOfBibitem
\bibitem[Wang \latin{et~al.}(1994)Wang, Schoenlein, Peteanu, Mathies, and
  Shank]{wang1994}
Wang,~Q.; Schoenlein,~R.~W.; Peteanu,~L.~A.; Mathies,~R.~A.; Shank,~C.~V.
  Vibrationally coherent photochemistry in the femtosecond primary event of
  vision. \emph{Science} \textbf{1994}, \emph{266}, 422--424\relax
\mciteBstWouldAddEndPuncttrue
\mciteSetBstMidEndSepPunct{\mcitedefaultmidpunct}
{\mcitedefaultendpunct}{\mcitedefaultseppunct}\relax
\EndOfBibitem
\bibitem[Kim \latin{et~al.}(2003)Kim, Tauber, and Mathies]{kim2003}
Kim,~J.~E.; Tauber,~M.~J.; Mathies,~R.~A. Analysis of the Mode-Specific
  Excited-State Energy Distribution and Wavelength-Dependent Photoreaction
  Quantum Yield in Rhodopsin. \emph{Biophys. J.} \textbf{2003}, \emph{84},
  2492--2501\relax
\mciteBstWouldAddEndPuncttrue
\mciteSetBstMidEndSepPunct{\mcitedefaultmidpunct}
{\mcitedefaultendpunct}{\mcitedefaultseppunct}\relax
\EndOfBibitem
\bibitem[Johnson \latin{et~al.}(2017)Johnson, Farag, Halpin, Morizumi,
  Prokhorenko, Knoester, Jansen, Ernst, and Miller]{johnson2017}
Johnson,~P.~J.; Farag,~M.~H.; Halpin,~A.; Morizumi,~T.; Prokhorenko,~V.~I.;
  Knoester,~J.; Jansen,~T.~L.; Ernst,~O.~P.; Miller,~R.~D. The primary
  photochemistry of vision occurs at the molecular speed limit. \emph{J. Phys.
  Chem. B} \textbf{2017}, \emph{121}, 4040--4047\relax
\mciteBstWouldAddEndPuncttrue
\mciteSetBstMidEndSepPunct{\mcitedefaultmidpunct}
{\mcitedefaultendpunct}{\mcitedefaultseppunct}\relax
\EndOfBibitem
\bibitem[Ao and Rammer(1991)Ao, and Rammer]{AoRammer1991PRB}
Ao,~P.; Rammer,~J. Quantum dynamics of a two-state system in a dissipative
  environment. \emph{Phys. Rev. B} \textbf{1991}, \emph{43}, 5397--5418\relax
\mciteBstWouldAddEndPuncttrue
\mciteSetBstMidEndSepPunct{\mcitedefaultmidpunct}
{\mcitedefaultendpunct}{\mcitedefaultseppunct}\relax
\EndOfBibitem
\bibitem[Kayanuma and Nakayama(1998)Kayanuma, and
  Nakayama]{KayanumaNakayama1998PRB}
Kayanuma,~Y.; Nakayama,~H. Nonadiabatic transition at a level crossing with
  dissipation. \emph{Phys. Rev. B} \textbf{1998}, \emph{57}, 13099--13112\relax
\mciteBstWouldAddEndPuncttrue
\mciteSetBstMidEndSepPunct{\mcitedefaultmidpunct}
{\mcitedefaultendpunct}{\mcitedefaultseppunct}\relax
\EndOfBibitem
\bibitem[Nalbach and Thorwart(2010)Nalbach, and
  Thorwart]{NalbachThorwart2010ChemPhys}
Nalbach,~P.; Thorwart,~M. Competition between relaxation and external driving
  in the dissipative Landau–Zener problem. \emph{Chem. Phys.} \textbf{2010},
  \emph{375}, 234--242\relax
\mciteBstWouldAddEndPuncttrue
\mciteSetBstMidEndSepPunct{\mcitedefaultmidpunct}
{\mcitedefaultendpunct}{\mcitedefaultseppunct}\relax
\EndOfBibitem
\bibitem[Dodin \latin{et~al.}(2014)Dodin, Garmon, Simine, and
  Segal]{Amro2014JCP}
Dodin,~A.; Garmon,~S.; Simine,~L.; Segal,~D. Landau-Zener transitions mediated
  by an environment: Population transfer and energy dissipation. \emph{J. Chem.
  Phys.} \textbf{2014}, \emph{140}, 124709\relax
\mciteBstWouldAddEndPuncttrue
\mciteSetBstMidEndSepPunct{\mcitedefaultmidpunct}
{\mcitedefaultendpunct}{\mcitedefaultseppunct}\relax
\EndOfBibitem
\bibitem[Novelli \latin{et~al.}(2015)Novelli, Belzig, and
  Nitzan]{NovelliBelzigNitzan2015NJP}
Novelli,~A.; Belzig,~W.; Nitzan,~A. Landau{\textendash}Zener evolution under
  weak measurement: manifestation of the Zeno effect under diabatic and
  adiabatic measurement protocols. \emph{New J. Phys.} \textbf{2015},
  \emph{17}, 013001\relax
\mciteBstWouldAddEndPuncttrue
\mciteSetBstMidEndSepPunct{\mcitedefaultmidpunct}
{\mcitedefaultendpunct}{\mcitedefaultseppunct}\relax
\EndOfBibitem
\bibitem[Farag \latin{et~al.}(2018)Farag, Jansen, and
  Knoester]{FaragJansenKnoester2018PCCP}
Farag,~M.~H.; Jansen,~T. L.~C.; Knoester,~J. The origin of absorptive features
  in the two-dimensional electronic spectra of rhodopsin. \emph{Phys. Chem.
  Chem. Phys.} \textbf{2018}, \emph{20}, 12746--12754\relax
\mciteBstWouldAddEndPuncttrue
\mciteSetBstMidEndSepPunct{\mcitedefaultmidpunct}
{\mcitedefaultendpunct}{\mcitedefaultseppunct}\relax
\EndOfBibitem
\bibitem[Marsili \latin{et~al.}(2019)Marsili, Farag, Yang, De~Vico, and
  Olivucci]{Olivucci2S3M2019JPCA}
Marsili,~E.; Farag,~M.~H.; Yang,~X.; De~Vico,~L.; Olivucci,~M. Two-State,
  Three-Mode Parametrization of the Force Field of a Retinal Chromophore Model.
  \emph{J. Phys. Chem. A} \textbf{2019}, \emph{123}, 1710--1719\relax
\mciteBstWouldAddEndPuncttrue
\mciteSetBstMidEndSepPunct{\mcitedefaultmidpunct}
{\mcitedefaultendpunct}{\mcitedefaultseppunct}\relax
\EndOfBibitem
\bibitem[Ottolenghi(1982)]{Ottolenghi1982}
Ottolenghi,~M. Molecular aspects of the photocycles of rhodopsin and
  bacteriorhodopsin: A comparative overview. \emph{Meth. Enzymol.}
  \textbf{1982}, \emph{88}, 470--491\relax
\mciteBstWouldAddEndPuncttrue
\mciteSetBstMidEndSepPunct{\mcitedefaultmidpunct}
{\mcitedefaultendpunct}{\mcitedefaultseppunct}\relax
\EndOfBibitem
\bibitem[Wand \latin{et~al.}(2013)Wand, Gdor, Zhu, Sheves, and
  Ruhman]{Sheves2013}
Wand,~A.; Gdor,~I.; Zhu,~J.; Sheves,~M.; Ruhman,~S. Shedding New Light on
  Retinal Protein Photochemistry. \emph{Annu. Rev. Phys. Chem.} \textbf{2013},
  \emph{64}, 437--458\relax
\mciteBstWouldAddEndPuncttrue
\mciteSetBstMidEndSepPunct{\mcitedefaultmidpunct}
{\mcitedefaultendpunct}{\mcitedefaultseppunct}\relax
\EndOfBibitem
\bibitem[Chung \latin{et~al.}(2012)Chung, Nanbu, and Ishida]{Ishida2012JPCB}
Chung,~W.~C.; Nanbu,~S.; Ishida,~T. QM/MM Trajectory Surface Hopping Approach
  to Photoisomerization of Rhodopsin and Isorhodopsin: The Origin of Faster and
  More Efficient Isomerization for Rhodopsin. \emph{J. Phys. Chem. B}
  \textbf{2012}, \emph{116}, 8009--8023\relax
\mciteBstWouldAddEndPuncttrue
\mciteSetBstMidEndSepPunct{\mcitedefaultmidpunct}
{\mcitedefaultendpunct}{\mcitedefaultseppunct}\relax
\EndOfBibitem
\bibitem[Gozem \latin{et~al.}(2017)Gozem, Luk, Schapiro, and
  Olivucci]{Olivucci2017ChemRev}
Gozem,~S.; Luk,~H.~L.; Schapiro,~I.; Olivucci,~M. Theory and Simulation of the
  Ultrafast Double-Bond Isomerization of Biological Chromophores.
  \emph{Chemical Reviews} \textbf{2017}, \emph{117}, 13502--13565\relax
\mciteBstWouldAddEndPuncttrue
\mciteSetBstMidEndSepPunct{\mcitedefaultmidpunct}
{\mcitedefaultendpunct}{\mcitedefaultseppunct}\relax
\EndOfBibitem
\bibitem[Schnedermann \latin{et~al.}(2018)Schnedermann, Yang, Liebel, Spillane,
  Lugtenburg, Fern{\'a}ndez, Valentini, Schapiro, Olivucci, Kukura,
  \latin{et~al.} others]{schnedermann2018}
Schnedermann,~C.; Yang,~X.; Liebel,~M.; Spillane,~K.; Lugtenburg,~J.;
  Fern{\'a}ndez,~I.; Valentini,~A.; Schapiro,~I.; Olivucci,~M.; Kukura,~P.
  \latin{et~al.}  Evidence for a vibrational phase-dependent isotope effect on
  the photochemistry of vision. \emph{Nat. Chem.} \textbf{2018}, \emph{10},
  449\relax
\mciteBstWouldAddEndPuncttrue
\mciteSetBstMidEndSepPunct{\mcitedefaultmidpunct}
{\mcitedefaultendpunct}{\mcitedefaultseppunct}\relax
\EndOfBibitem
\bibitem[Zhu and Nakamura(1992)Zhu, and Nakamura]{ZhuNakamura1992JCP2}
Zhu,~C.; Nakamura,~H. The two‐state linear curve crossing problems revisited.
  II. Analytical approximations for the Stokes constant and scattering matrix:
  The Landau–Zener case. \emph{J. Chem. Phys.} \textbf{1992}, \emph{97},
  8497--8514\relax
\mciteBstWouldAddEndPuncttrue
\mciteSetBstMidEndSepPunct{\mcitedefaultmidpunct}
{\mcitedefaultendpunct}{\mcitedefaultseppunct}\relax
\EndOfBibitem
\bibitem[Tully(2012)]{TullyPerspective2012JCP}
Tully,~J.~C. Perspective: Nonadiabatic dynamics theory. \emph{J. Chem. Phys.}
  \textbf{2012}, \emph{137}, 22A301\relax
\mciteBstWouldAddEndPuncttrue
\mciteSetBstMidEndSepPunct{\mcitedefaultmidpunct}
{\mcitedefaultendpunct}{\mcitedefaultseppunct}\relax
\EndOfBibitem
\bibitem[Suchan \latin{et~al.}(2020)Suchan, Janoš, and
  Slavíček]{Slavicek2020JCTC}
Suchan,~J.; Janoš,~J.; Slavíček,~P. Pragmatic Approach to Photodynamics:
  Mixed Landau–Zener Surface Hopping with Intersystem Crossing. \emph{J.
  Chem. Theory Comput.} \textbf{2020}, \emph{16}, 5809--5820\relax
\mciteBstWouldAddEndPuncttrue
\mciteSetBstMidEndSepPunct{\mcitedefaultmidpunct}
{\mcitedefaultendpunct}{\mcitedefaultseppunct}\relax
\EndOfBibitem
\bibitem[El-Tahawy \latin{et~al.}(2018)El-Tahawy, Nenov, Weingart, Olivucci,
  and Garavelli]{ElTahawy2018JPCL}
El-Tahawy,~M. M.~T.; Nenov,~A.; Weingart,~O.; Olivucci,~M.; Garavelli,~M.
  Relationship between Excited State Lifetime and Isomerization Quantum Yield
  in Animal Rhodopsins: Beyond the One-Dimensional Landau–Zener Model.
  \emph{J. Phys. Chem. Lett.} \textbf{2018}, \emph{9}, 3315--3322\relax
\mciteBstWouldAddEndPuncttrue
\mciteSetBstMidEndSepPunct{\mcitedefaultmidpunct}
{\mcitedefaultendpunct}{\mcitedefaultseppunct}\relax
\EndOfBibitem
\bibitem[Hoki and Brumer(2011)Hoki, and Brumer]{Hoki2011}
Hoki,~K.; Brumer,~P. Excitation of Biomolecules by Coherent vs. Incoherent
  Light: Model Rhodopsin Photoisomerization. \emph{Procedia Chemistry}
  \textbf{2011}, \emph{3}, 122 -- 131, 22nd Solvay Conference on
  Chemistry\relax
\mciteBstWouldAddEndPuncttrue
\mciteSetBstMidEndSepPunct{\mcitedefaultmidpunct}
{\mcitedefaultendpunct}{\mcitedefaultseppunct}\relax
\EndOfBibitem
\bibitem[Gruebele and Wolynes(2004)Gruebele, and
  Wolynes]{GruebeleWolynes2004AccChemRes}
Gruebele,~M.; Wolynes,~P.~G. Vibrational Energy Flow and Chemical Reactions.
  \emph{Acc. Chem. Res.} \textbf{2004}, \emph{37}, 261--267\relax
\mciteBstWouldAddEndPuncttrue
\mciteSetBstMidEndSepPunct{\mcitedefaultmidpunct}
{\mcitedefaultendpunct}{\mcitedefaultseppunct}\relax
\EndOfBibitem
\bibitem[Xie and Domcke(2017)Xie, and Domcke]{XieDomcke2017JCP}
Xie,~W.; Domcke,~W. Accuracy of trajectory surface-hopping methods: Test for a
  two-dimensional model of the photodissociation of phenol. \emph{J. Chem.
  Phys.} \textbf{2017}, \emph{147}, 184114\relax
\mciteBstWouldAddEndPuncttrue
\mciteSetBstMidEndSepPunct{\mcitedefaultmidpunct}
{\mcitedefaultendpunct}{\mcitedefaultseppunct}\relax
\EndOfBibitem
\bibitem[Balzer and Stock(2005)Balzer, and Stock]{balzer2005}
Balzer,~B.; Stock,~G. Modeling of decoherence and dissipation in nonadiabatic
  photoreactions by an effective-scaling nonsecular Redfield algorithm.
  \emph{Chem. Phys.} \textbf{2005}, \emph{310}, 33--41\relax
\mciteBstWouldAddEndPuncttrue
\mciteSetBstMidEndSepPunct{\mcitedefaultmidpunct}
{\mcitedefaultendpunct}{\mcitedefaultseppunct}\relax
\EndOfBibitem
\bibitem[Marston and Balint-Kurti(1989)Marston, and
  Balint-Kurti]{MarstonBalintKurti1989JCP}
Marston,~C.~C.; Balint-Kurti,~G.~G. The Fourier grid Hamiltonian method for
  bound state eigenvalues and eigenfunctions. \emph{J. Chem. Phys.}
  \textbf{1989}, \emph{91}, 3571--3576\relax
\mciteBstWouldAddEndPuncttrue
\mciteSetBstMidEndSepPunct{\mcitedefaultmidpunct}
{\mcitedefaultendpunct}{\mcitedefaultseppunct}\relax
\EndOfBibitem
\bibitem[Chuang and Brumer(2021)Chuang, and Brumer]{RetinalChaos1}
Chuang,~C.; Brumer,~P. Extreme Parametric Sensitivity in the Steady-State
  Photoisomerization of Two-Dimensional Rhodopsin Model. \emph{J. Phys. Chem.
  Lett.} \textbf{2021}, \emph{16}, 3618\relax
\mciteBstWouldAddEndPuncttrue
\mciteSetBstMidEndSepPunct{\mcitedefaultmidpunct}
{\mcitedefaultendpunct}{\mcitedefaultseppunct}\relax
\EndOfBibitem
\bibitem[Hahn and Stock(2002)Hahn, and Stock]{hahn2002}
Hahn,~S.; Stock,~G. Ultrafast cis-trans photoswitching: A model study. \emph{J.
  Chem. Phys.} \textbf{2002}, \emph{116}, 1085--1091\relax
\mciteBstWouldAddEndPuncttrue
\mciteSetBstMidEndSepPunct{\mcitedefaultmidpunct}
{\mcitedefaultendpunct}{\mcitedefaultseppunct}\relax
\EndOfBibitem
\bibitem[Eyring \latin{et~al.}(1980)Eyring, Curry, Mathies, Fransen, Palings,
  and Lugtenburg]{Eyring1980}
Eyring,~G.; Curry,~B.; Mathies,~R.~A.; Fransen,~R.; Palings,~I.; Lugtenburg,~J.
  Interpretation of the Resonance Raman Spectrum of Bathorhodopsin Based on
  Visual Pigment Analogues. \emph{Biochem.} \textbf{1980}, \emph{19},
  2410--2418\relax
\mciteBstWouldAddEndPuncttrue
\mciteSetBstMidEndSepPunct{\mcitedefaultmidpunct}
{\mcitedefaultendpunct}{\mcitedefaultseppunct}\relax
\EndOfBibitem
\bibitem[Ishizaki and Fleming({2009})Ishizaki, and Fleming]{Ishizaki2009}
Ishizaki,~A.; Fleming,~G.~R. {On the adequacy of the Redfield equation and
  related approaches to the study of quantum dynamics in electronic energy
  transfer}. \emph{{J. Chem. Phys.}} \textbf{{2009}}, \emph{{130}},
  {234110}\relax
\mciteBstWouldAddEndPuncttrue
\mciteSetBstMidEndSepPunct{\mcitedefaultmidpunct}
{\mcitedefaultendpunct}{\mcitedefaultseppunct}\relax
\EndOfBibitem
\bibitem[Prokhorenko \latin{et~al.}(2006)Prokhorenko, Nagy, Waschuk, Brown,
  Birge, and Miller]{prokhorenko2006}
Prokhorenko,~V.~I.; Nagy,~A.~M.; Waschuk,~S.~A.; Brown,~L.~S.; Birge,~R.~R.;
  Miller,~R.~D. Coherent control of retinal isomerization in bacteriorhodopsin.
  \emph{Science} \textbf{2006}, \emph{313}, 1257--1261\relax
\mciteBstWouldAddEndPuncttrue
\mciteSetBstMidEndSepPunct{\mcitedefaultmidpunct}
{\mcitedefaultendpunct}{\mcitedefaultseppunct}\relax
\EndOfBibitem
\bibitem[Köuppel \latin{et~al.}(1984)Köuppel, Domcke, and
  Cederbaum]{KoppelDomckeCederbaum1984ACP}
Köuppel,~H.; Domcke,~W.; Cederbaum,~L.~S. \emph{Adv. Chem. Phys.}; John Wiley
  and Sons, Ltd, 1984; pp 59--246\relax
\mciteBstWouldAddEndPuncttrue
\mciteSetBstMidEndSepPunct{\mcitedefaultmidpunct}
{\mcitedefaultendpunct}{\mcitedefaultseppunct}\relax
\EndOfBibitem
\bibitem[Heller(1990)]{Heller1990}
Heller,~E. Mode mixing and chaos induced by potential surface crossings.
  \emph{J. Chem. Phys.} \textbf{1990}, \emph{92}, 1718\relax
\mciteBstWouldAddEndPuncttrue
\mciteSetBstMidEndSepPunct{\mcitedefaultmidpunct}
{\mcitedefaultendpunct}{\mcitedefaultseppunct}\relax
\EndOfBibitem
\bibitem[Leitner \latin{et~al.}({1996})Leitner, K{\"{o}}ppel, and
  Cederbaum]{LeitnerCederbaum1996}
Leitner,~D.~M.; K{\"{o}}ppel,~H.; Cederbaum,~L.~S. {Statistical properties of
  molecular spectra and molecular dynamics: Analysis of their correspondence in
  NO2 and C2H+4}. \emph{{J. Chem. Phys.}} \textbf{{1996}}, \emph{{104}},
  {434}\relax
\mciteBstWouldAddEndPuncttrue
\mciteSetBstMidEndSepPunct{\mcitedefaultmidpunct}
{\mcitedefaultendpunct}{\mcitedefaultseppunct}\relax
\EndOfBibitem
\bibitem[Chen \latin{et~al.}(2016)Chen, Gelin, Chernyak, Domcke, and
  Zhao]{DomckeZhao2016}
Chen,~L.; Gelin,~M.~F.; Chernyak,~V.~Y.; Domcke,~W.; Zhao,~Y. Dissipative
  dynamics at conical intersections: simulations with the hierarchy equations
  of motion method. \emph{Faraday Discuss.} \textbf{2016}, \emph{194},
  61--80\relax
\mciteBstWouldAddEndPuncttrue
\mciteSetBstMidEndSepPunct{\mcitedefaultmidpunct}
{\mcitedefaultendpunct}{\mcitedefaultseppunct}\relax
\EndOfBibitem
\bibitem[Chung \latin{et~al.}(2007)Chung, Witkoskie, Zimmer, Cao, and
  Bawendi]{CaoBawendi2007PRB}
Chung,~I.; Witkoskie,~J.~B.; Zimmer,~J.~P.; Cao,~J.; Bawendi,~M.~G. Extracting
  the number of quantum dots in a microenvironment from ensemble fluorescence
  intensity fluctuations. \emph{Phys. Rev. B} \textbf{2007}, \emph{75},
  045311\relax
\mciteBstWouldAddEndPuncttrue
\mciteSetBstMidEndSepPunct{\mcitedefaultmidpunct}
{\mcitedefaultendpunct}{\mcitedefaultseppunct}\relax
\EndOfBibitem
\bibitem[Landry and Subotnik(2011)Landry, and Subotnik]{LandrySubotnik2011JCP}
Landry,~B.~R.; Subotnik,~J.~E. Communication: Standard surface hopping predicts
  incorrect scaling for Marcus’ golden-rule rate: The decoherence problem
  cannot be ignored. \emph{J. Chem. Phys.} \textbf{2011}, \emph{135},
  191101\relax
\mciteBstWouldAddEndPuncttrue
\mciteSetBstMidEndSepPunct{\mcitedefaultmidpunct}
{\mcitedefaultendpunct}{\mcitedefaultseppunct}\relax
\EndOfBibitem
\bibitem[Ashhab(2014)]{Ashhab2014PRA}
Ashhab,~S. Landau-Zener transitions in a two-level system coupled to a
  finite-temperature harmonic oscillator. \emph{Phys. Rev. A} \textbf{2014},
  \emph{90}, 062120\relax
\mciteBstWouldAddEndPuncttrue
\mciteSetBstMidEndSepPunct{\mcitedefaultmidpunct}
{\mcitedefaultendpunct}{\mcitedefaultseppunct}\relax
\EndOfBibitem
\end{mcitethebibliography}
\end{document}